\documentclass[times,twocolumn,final]{elsarticle}

\usepackage{arxiv}
\usepackage{graphicx}%
\usepackage{multirow}%
\usepackage{amsmath,amssymb,amsfonts}%

\usepackage{amsthm}%
\usepackage{mathrsfs}%
\usepackage[title]{appendix}%
\usepackage{xcolor}%
\usepackage[utf8]{inputenc}
\usepackage{booktabs}%
\usepackage{algpseudocode,algorithm}
\MakeRobust{\Call}
\usepackage{pifont}
\usepackage{adjustbox}
\usepackage{graphicx}
\usepackage{float}
\usepackage{placeins}
\usepackage{stfloats}
\usepackage{fancyhdr}
\usepackage[hyphens]{url}
\usepackage{hyperref}
\usepackage{caption}
\usepackage{qrcode}
\usepackage{longtable}
\usepackage{pdfpages}

\usepackage{subcaption}
\usepackage{booktabs,tabularx,ragged2e,array,multirow,graphicx}
\newcolumntype{Y}{>{\RaggedRight\arraybackslash}X}
\renewcommand{\arraystretch}{1.15}

\fancypagestyle{firstpagestyle}{
    \fancyhf{} 
    \fancyhead{} 
    \fancyfoot{} 
}

\fancypagestyle{default}{
    \fancyhf{}
    \fancyhead[R]{\thepage} 
}

\begin{document}

\title{Optimizing Human-Machine Interface for Real-Time AI Support in the Operating Room: the CVS Copilot}

\author[1]{Lorenzo \snm{Arboit}\corref{corresp}\fnref{first}}
\cortext[corresp]{Corresponding author: \texttt{larboit@unistra.fr}}
\author[1]{Nicolas \snm{Chanel}\fnref{first}}
\author[1]{Aditya \snm{Murali}}
\author[2,3]{Pietro \snm{Mascagni}\fnref{equal}}
\author[1,2]{Nicolas \snm{Padoy}\fnref{equal}}

\address[1]{University of Strasbourg, CNRS, INSERM, ICube, UMR7357, Strasbourg, France}
\address[2]{IHU Strasbourg, Strasbourg, France}
\address[3]{Fondazione Policlinico Universitario Agostino Gemelli IRCCS, Rome, Italy}

\fntext[first]{These authors contributed equally and share first co-authorship}
\fntext[equal]{These authors contributed equally and share last co-authorship }

\received{XXX}
\finalform{XXX}
\accepted{XXX}
\availableonline{XXX}
\communicated{XXX}

\begin{abstract}
Artificial intelligence (AI) systems for automated Critical View of Safety (CVS) assessment in laparoscopic cholecystectomy are nearing clinical translation. Beyond algorithmic performance, clinical safety and effectiveness depend on the quality of the human–machine interface (HMI). This work examines how AI-generated predictions should be presented and controlled intraoperatively. Seventeen surgeons, including residents, attending surgeons, and professors, took part in a mixed-methods, user-centered design study to optimize an intraoperative HMI for AI-assisted safe laparoscopic cholecystectomy. Interviews explored interaction modalities, timing of assistance, visualization strategies, and control mechanisms across surgical roles, and were analyzed using reflexive thematic analysis and human-factors heuristics. Most surgeons (16/17) supported the use of AI for intraoperative decision support while rejecting autonomous decision-making. Attendings preferred minimal AI feedback at decisive moments (13/14), whereas residents favored optional guidance (3/3) with confidence indicators and on-demand anatomical overlays. Across interviews, surgeons consistently prioritized visual, surgeon-controlled, minimally intrusive displays, with the strongest support for a minimal overlay (16/17) and on-demand anatomical segmentation (13/17). Recurrent concerns included persistent overlays, haptic feedback, and numeric confidence displays, although these were not uniformly raised across the cohort. These findings informed the design of CVS Copilot, a surgeon-controlled, role-adaptive HMI that provides AI-based CVS assessment with minimal default visualization and optional overlays.
\end{abstract}

\maketitle
\thispagestyle{firstpagestyle}

\section{Introduction}
\label{sec:introduction}

Advances in computational methods for endoscopic video analysis have enabled large-scale, data-driven research in surgery, with laparoscopic cholecystectomy (LC) emerging as one of the most widely studied procedures~\cite{mascagni_computer_2022}. This is largely due to the high global volume of LC, the relatively standardized nature of its operative workflow, and the vast consensus on a visually defined best practice. In fact, a key safety milestone within this procedure is the achievement of the Critical View of Safety (CVS), consisting in the clear visualization of surgical landmarks to prevent major bile duct injuries. First described by Strasberg in 1995~\cite{strasberg_analysis_1995}, the CVS is now recommended by major surgical societies~\cite{brunt_safe_2020}.

Several artificial intelligence (AI)-based computer vision models have been proposed to automate the assessment of CVS in LC. Early approaches such as DeepCVS~\cite{mascagni_artificial_2022} combined hepatocystic anatomy segmentation with binary classification of the three criteria defining the CVS. Subsequent work introduces more advanced representations, including graph-based frameworks that model anatomical relationships to predict CVS achievement~\cite{murali_latent_2024}, as well as spatiotemporal architectures such as SwinCVS~\cite{nowak_swincvs_2025} trained on larger annotated datasets to capture the dynamic evolution of dissection. More recently, the SAGES CVS Challenge~\cite{alapatt_sages_2025} established the first large-scale public benchmark for CVS assessment by providing a multi-institutional dataset of expert-annotated laparoscopic cholecystectomy videos to evaluate model accuracy and generalizability across centers.

These developments mark an important step toward the clinical translation of AI-based intraoperative guidance systems. Since the early technical proof-of-concept demonstrations of real-time intraoperative AI guidance in 2021~\cite{mascagni_early-stage_2024}, translational studies are fast evolving into clinical trials currently underway~\cite{madani_evaluating_2025,ihu_strasbourg_cvs-notifier_2025,sun_yat-sen_memorial_hospital_of_sun_yat-sen_university_lc-smart_2024}. However, successful integration of AI into the operating room depends not only on predictive performance but also on how model outputs are communicated to surgeons during the procedure~\cite{topol_high-performance_2019}.

In high-stakes environments such as the operating room, poorly designed interfaces may contribute to sensory overload, disrupt workflow, and increase cognitive load, potentially undermining clinical utility and safety. Evidence from other clinical domains supports this concern. In intensive care, excessive and poorly prioritized alarms have been shown to desensitize healthcare professionals and reduce vigilance~\cite{dursun_ergezen_nurses_2020}, while in diagnostic imaging, AI tools with suboptimal user interfaces have struggled with adoption despite proven improvements in detection rates~\cite{gaube_as_2021}. Similarly, studies of intraoperative visualization technologies, such as near-infrared fluorescence and augmented reality overlays, highlight that surgeons consistently demand minimalism, glanceability, and control over when and how information appears~\cite{diana_enhanced-reality_2014,meola_augmented_2017,vavra_recent_2017}.

So far, relatively little attention has been devoted to understanding how intraoperative AI predictions should be effectively presented to surgeons during surgery. To fill this gap, this work adopts a user-centered design (UCD) framework to study and develop an optimized interface for real-time AI-based CVS assessment in the operating room, the CVS Copilot.

\section{Results}
\label{sec:results}

Seventeen surgeons participated in semi-structured interviews exploring user experience, user interface preferences, and mock-up evaluation for an AI system supporting intraoperative CVS assessment. Participants included 3 residents, 11 attending surgeons, and 3 professors from institutions across Europe and the United States; detailed surgeons’ characteristics are presented in Table~\ref{tab:table1}.

\begin{table}[h]
{\small
\setlength{\tabcolsep}{3pt}
\renewcommand{\arraystretch}{1.1}
\centering
\caption{Participant characteristics of the interviewed surgeons (n = 17).}
\label{tab:table1}
\begin{tabular}{lll}
\toprule
\textbf{Characteristic} & \textbf{n} & \textbf{\%} \\
\midrule
\textbf{Gender} & & \\
Male & 14 & 82 \\
Female & 3 & 18 \\
\addlinespace
\textbf{Country} & & \\
United States & 1 & 6 \\
France & 1 & 6 \\
Germany & 1 & 6 \\
Spain & 2 & 12 \\
Italy & 12 & 70 \\
\addlinespace
\textbf{Role} & & \\
Senior Residents & 3 & 18 \\
Attending Surgeons & 11 & 64 \\
Professors & 3 & 18 \\
\addlinespace
\textbf{Practice setting} & & \\
Academic & 17 & 100 \\
\addlinespace
\textbf{Surgical experience (years)} & & \\
$<$ 5 & 3 & 18 \\
5--10 & 5 & 29 \\
10--20 & 7 & 41 \\
$>$ 20 & 2 & 12 \\
\addlinespace
\textbf{Prior exposure to surgical AI} & & \\
Involved in SDS projects & 17 & 100 \\
\bottomrule
\end{tabular}
}
\end{table}

Interviews were conducted remotely using a semi-structured format based on predefined scenarios and interface mock-ups, while allowing participants to elaborate freely on workflow integration, interaction preferences, and safety considerations. All sessions were audio-video recorded, transcribed, and iteratively analyzed throughout the study process.
Interviews produced 496 coded excerpts, corresponding to distinct transcript segments assigned thematic codes, and thematic saturation was reached after thirteen interviews. A structured thematic analysis identified three overarching categories, namely User Experience (UX), User Interface (UI), and Interpretability/Interaction, each comprising multiple themes (i.e., broader recurring concepts capturing shared surgeon perspectives; 10 in total), codes (i.e., granular labels assigned to specific ideas or interaction preferences; 31 in total), and representative quotations (Table~\ref{tab:table2}). These categories reflected distinct but interrelated components of how surgeons envision intraoperative assistance for an AI system. Surgeons did not provide feedback on the final themes.

\begin{table*}[p]
\centering
\caption{Surgeon-derived themes and design requirements for an AI-based CVS Copilot interface.}
\label{tab:table2}
{\scriptsize
\setlength{\tabcolsep}{3pt}
\renewcommand{\arraystretch}{1.05}
\begin{adjustbox}{max totalsize={\textwidth}{0.92\textheight},center}
\begin{tabular}{>{\centering\arraybackslash}m{0.05\textwidth} p{0.15\textwidth} p{0.16\textwidth} p{0.22\textwidth} p{0.34\textwidth}}
	\textbf{Category} & \textbf{Theme} & \textbf{Code} & \textbf{Definition} & \textbf{Representative Quote} \\
\midrule
\multirow{13}{*}{\rotatebox[origin=c]{90}{\textbf{User Experience (UX)}}} & Attitude Toward AI & AI as expert support & AI viewed as an additional expert improving safety & ``It’s like having a senior surgeon always with you.'' (surgeon n.\ 15) \\
& & Need for evidence of safety and effectiveness & AI accepted only if non-intrusive and helpful & ``AI is useful only if it lowers the cognitive load, not if it becomes another noise.'' (surgeon n.\ 14) \\
& & Legal/privacy concerns & Concern about documentation and medical-legal exposure & ``People fear being recorded if they make mistakes.'' (surgeon n.\ 7) \\
\addlinespace
& OR Integration \& Setup & Zero-friction workflow & AI must not disrupt the operative flow & ``Everything bulky or complicated will not be used.'' (surgeon n.\ 14) \\
& & Pre-operative readiness & System must be ready before surgeon enters & ``Setup must be ready before the surgeon enters the OR.'' (surgeon n.\ 7) \\
& & Integration in tower & System must be embedded in existing laparoscopic tower & ``The best would be a simple software connected to the laparoscopic unit.'' (surgeon n.\ 2) \\
\addlinespace
& Control \& Activation & Manual toggle required & Surgeon has manual control on when AI appears & ``I want to manually control when to show it.'' (surgeon n.\ 5) \\
& & Camera-head button & Button on the camera to activate the Copilot & ``A button on the camera is best, no extra hardware.'' (surgeon n.\ 4) \\
& & Voice control & Voice to activate the Copilot & ``Voice control should be the best. I mean, it’s like Alexa.'' (surgeon n.\ 9) \\
& & Conditional automation & Auto-activation only in specific moments & ``If it knows I’m inserting the clipper, it can appear.'' (surgeon n.\ 15) \\
\addlinespace
& Timing of AI Assistance & Final-stage confirmation & Green-light before clipping and cutting & ``Just before clipping, give me the green light.'' (surgeon n.\ 8) \\
& & On-demand assistance & Surgeon queries AI when needed & ``Let me call the AI when I need it.'' (surgeon n.\ 5) \\
& & Early guidance for trainees & Residents want early feedback & ``Residents may need more information earlier.'' (surgeon n.\ 14) \\
\addlinespace
\multirow{10}{*}{\rotatebox[origin=c]{90}{\textbf{User Interface (UI)}}} & Information Modality & Visual dominance & Visual information preferred & ``Visual only, audio is for emergencies.'' (surgeon n.\ 6) \\
& & Audio only for danger & Audio should be reserved for alerts & ``Audio is for danger, not routine guidance.'' (surgeon n.\ 6) \\
& & Reject haptics & No vibration/blocking of tools & ``Never touch my instruments.'' (surgeon n.\ 4) \\
\addlinespace
& Visual Design Preferences & Minimal overlay & Small, corner-mounted icon triad & ``I want something glanceable that doesn’t cover anatomy.'' (surgeon n.\ 14) \\
& & Color-coded criteria & Green/amber/red states for CVS & ``Three colors are sufficient.'' (surgeon n.\ 1) \\
& & Dashboard on-demand & Expanded UI only when requested & ``Give me the simple view by default; let me expand when I need more.'' (surgeon n.\ 2) \\
& & Reject persistent overlays & No constant segmentation on screen & ``Overlays clutter the field unless I ask for them.'' (surgeon n.\ 9) \\
& & Transient segmentation & Only when explanation is needed & ``Highlight the artery and duct only when I ask.'' (surgeon n.\ 11) \\
\addlinespace
& Confidence Communication & Visual confidence & Shading or color to indicate certainty & ``A probability-like indication is useful, not numbers.'' (surgeon n.\ 12) \\
& & Avoid numeric precision & Avoid \% values to reduce overtrust & ``Numbers make you believe it too much.'' (surgeon n.\ 17) \\
\addlinespace
\multirow{8}{*}{\rotatebox[origin=c]{90}{\textbf{Interaction \& Interpretability}}} & Handling Disagreement / Mismatch & Need for justification & AI should explain why CVS is not achieved & ``If it’s wrong, tell me why it’s wrong.'' (surgeon n.\ 4) \\
& & Highlight problematic area & AI should show missing dissection & ``Show me exactly where dissection is missing.'' (surgeon n.\ 17) \\
& & Layered interpretability & Explanations available only when requested & ``It has to be a nice suggestion, not an order.'' (surgeon n.\ 12) \\
\addlinespace
& Clinical Exceptions & Recognizing impossible CVS & System must detect when CVS is anatomically impossible & ``Sometimes CVS will never be achieved; you must allow me to proceed safely.'' (surgeon n.\ 15) \\
& & Avoid penalizing surgeons & AI should not be rigid in difficult cases & ``If I burned the artery, of course I won’t achieve the first criterion.'' (surgeon n.\ 15) \\
\addlinespace
& Training \& Education & Resident-centered detail & Trainees need more detail & ``Residents may benefit from audio descriptions.'' (surgeon n.\ 14) \\
& & Expert-centered minimalism & Experts want condensed outputs & ``Keep it simple.'' (surgeon n.\ 12) \\
& & Adaptive interface & Different layers for different roles & ``One interface won’t work for experts and residents alike.'' (surgeon n.\ 14) \\
\bottomrule
\end{tabular}
\end{adjustbox}
}
\end{table*}

\subsection{User Experience}

All but one of the interviewed surgeons (94.1\%) explicitly endorsed the concept of intraoperative AI as expert support, often describing it as an additional layer of safety or an ``expert colleague'' in the room. Five surgeons (29.4\%) reported that acceptance was conditional on the system being non-intrusive and reducing cognitive load. Six surgeons (35.3\%) raised specific legal or privacy concerns regarding data recording, noting that colleagues might fear ``being recorded if they make mistakes''.

A requirement for a zero-friction workflow was reported by 52.9\% (9/17) of participants. Surgeons insisted that the AI system must not disrupt the operative flow or require complex setup procedures. To achieve this, 47.1\% (8/17) specifically requested that the system be physically embedded or integrated directly into the existing laparoscopic tower to avoid external hardware clutter, with one surgeon noting that ``everything bulky or complicated will not be used''.

With respect to system control, most surgeons (76.5\%, 13/17) required a manual toggle to explicitly control the visibility of AI outputs, rather than automatically triggered overlays. The preferred mechanisms for this control varied, with 35.3\% (6/17) specifically suggesting a button on the camera head as the ideal mechanism for sterile, immediate access. Opinions on voice control differed across participants: while 52.9\% (9/17) of surgeons preferred hands-free voice interaction, 35.3\% (6/17) actively rejected it due to concerns over reliability in a noisy environment, such as the operating room (OR), or privacy issues. One surgeon opposing the technology argued that ``the OR is too chaotic for voice commands'', whereas proponents viewed it as ``extremely attractive'' and ``definitely the best option''.

Regarding the timing of AI feedback, 94.1\% (16/17) of surgeons preferred an on-demand assistance model, allowing them to query the AI only when needed rather than receiving constant unsolicited feedback. Nevertheless, 70.6\% (12/17) agreed that a ``final-stage confirmation'' (a ``green-light'' check immediately before division of the cystic duct and cystic artery) was the most critical moment for AI intervention to ensure safety.

\subsection{User Interface}

All surgeons preferred visual information over other sensory feedback. The majority of surgeons, 58.8\% (10/17), rejected routine audio feedback. Only 29.4\% (5/17) accepted audio, and strictly for danger alerts, noting that ``audio is for danger, not routine guidance''. Similarly, haptic feedback, such as tool vibration, was rejected by 41.2\% (7/17) of surgeons, who viewed it as unsafe or intrusive, with one surgeon remarking that the system should ``never touch my instruments''.

Regarding visual density, a total of 94.1\% (16/17) surgeons supported a minimal overlay, such as small corner icons or a traffic light system, that does not obstruct the anatomy. While persistent segmentation masks were generally considered clutter, 76.5\% (13/17) of surgeons supported transient on-demand segmentation (overlays that appear only when prompted) to clarify ambiguous anatomy. Concerning confidence communication, 41.2\% (7/17) of surgeons advised against displaying precise numeric percentages, fearing they might induce over-trust or confusion compared to binary signals (``numbers make you believe it too much''). In fact, 29.4\% (5/17) preferred visual analogues, such as color shading, to represent confidence.

\subsection{Interaction and Interpretability}

With regards to explainability, 64.7\% (11/17) of surgeons stated that binary outputs of ``safe'' or ``unsafe'' were insufficient when AI assessments conflicted with their judgment. They insisted that if the AI flags a CVS criterion as incomplete, it must provide a justification, with one attending stating, ``If it’s wrong, tell me why it’s wrong''. A total of 35.3\% (6/17) of surgeons requested that the system must highlight the specific problematic area directly on the screen to guide further dissection. Furthermore, a subset of 64.7\% (11/17) surgeons strongly emphasized that the AI must recognize ``impossible CVS'' scenarios, such as severe inflammation, to avoid penalizing surgeons for performing a safe sub-total cholecystectomy. As one surgeon summarized, ``Sometimes CVS will never be achieved; you must allow me to proceed safely''.

\begin{figure}[tbp]
    \centering
    \includegraphics[width=1\linewidth]{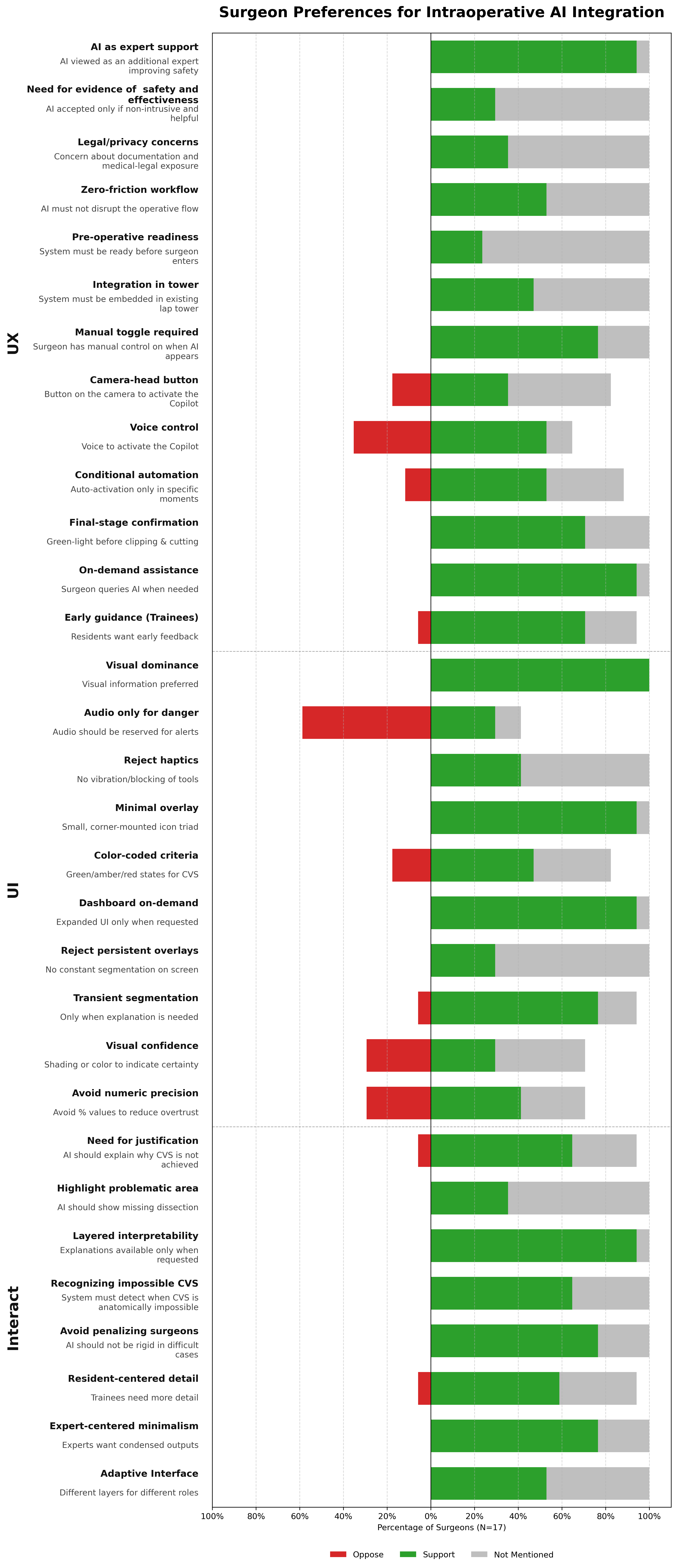}
    \caption{\textbf{Surgeon preferences for intraoperative AI integration in laparoscopic cholecystectomy.} \\ Stacked horizontal bars show the proportion of surgeons (n = 17) who supported (green), opposed (red), or did not mention (gray) each design feature identified through thematic analysis. The 31 codes are reported in bold, with their description following.}
    \label{fig:fig1}
\end{figure}

\subsection{Seniority-Based Differences}

Comparison between senior (Attendings \& Professors, n=14) and junior (Residents, n=3) surgeons revealed distinct preferences regarding the level of detail and guidance required. Junior surgeons were unanimous (100\%, 3/3) in their preference for interfaces that provide early guidance and educational feedback throughout the dissection phase. They viewed the AI as a potential teaching tool that could offer more intense guidance when attending surgeons are less communicative. In contrast, 92.9\% (13/14) of senior surgeons were significantly more inclined toward expert-centered minimalism. While they supported the concept of AI assistance, they preferred it to remain silent and invisible for the majority of the procedure, interacting only at the final confirmation stage. Senior surgeons explicitly stated a desire for a system that is ``simple'' and ``not bothersome,'' whereas junior surgeons were more open to continuous, layered interpretability to aid their learning curve. When prompted to adopt a resident's perspective, 50.0\% (7/14) of senior surgeons reported similar preferences, with only one surgeon suggesting the opposite.

Figure~\ref{fig:fig1} summarizes surgeons’ preferences across user experience, interface, and interaction dimensions, highlighting areas of consensus, rejection, and role-dependent divergence.

\subsection{Mock-Up Evaluation and CVS Copilot Design}

Predefined interface mock-ups were introduced during the final phase of each interview as structured probes to elicit preferences across alternative design configurations. The evaluation of these mock-ups, combined with earlier responses, directly informed the final design of the CVS Copilot, resulting in a dual-mode interface that addresses the conflicting needs for minimalism and detail. The first mode, serving as the default view, is the Minimal Interface, featuring a non-intrusive, corner-mounted status indicator. This design was accepted by 94.1\% (16/17) of surgeons, and it provides monitoring without obscuring the surgical field. The second stage is the Full Dashboard, available only on-demand. This expandable view provides confidence metrics, frame-by-frame analysis, and anatomy overlays. Other features highlighted during the discussions, such as text explanations, inability to safely achieve CVS, and dissection guides, were not introduced, as the models considered in this study do not yet provide sufficiently accurate and reliable predictions to support them. The final interface mockups are available in Figure~\ref{fig:fig2}A/B, and a video of the interface is available in Supplementary Video~1.
To preserve workflow integration, the system operates continuously in the background while visual information follows a controlled-disclosure logic: surgeons can manually activate both interfaces through a dedicated camera-head button, and the system can optionally smartly trigger the Minimal Interface at safety-critical moments such as clipper insertion or achievement of individual CVS criteria (Figure~\ref{fig:fig2}C).

\begin{figure*}[t]
    \centering
    \includegraphics[width=1\linewidth]{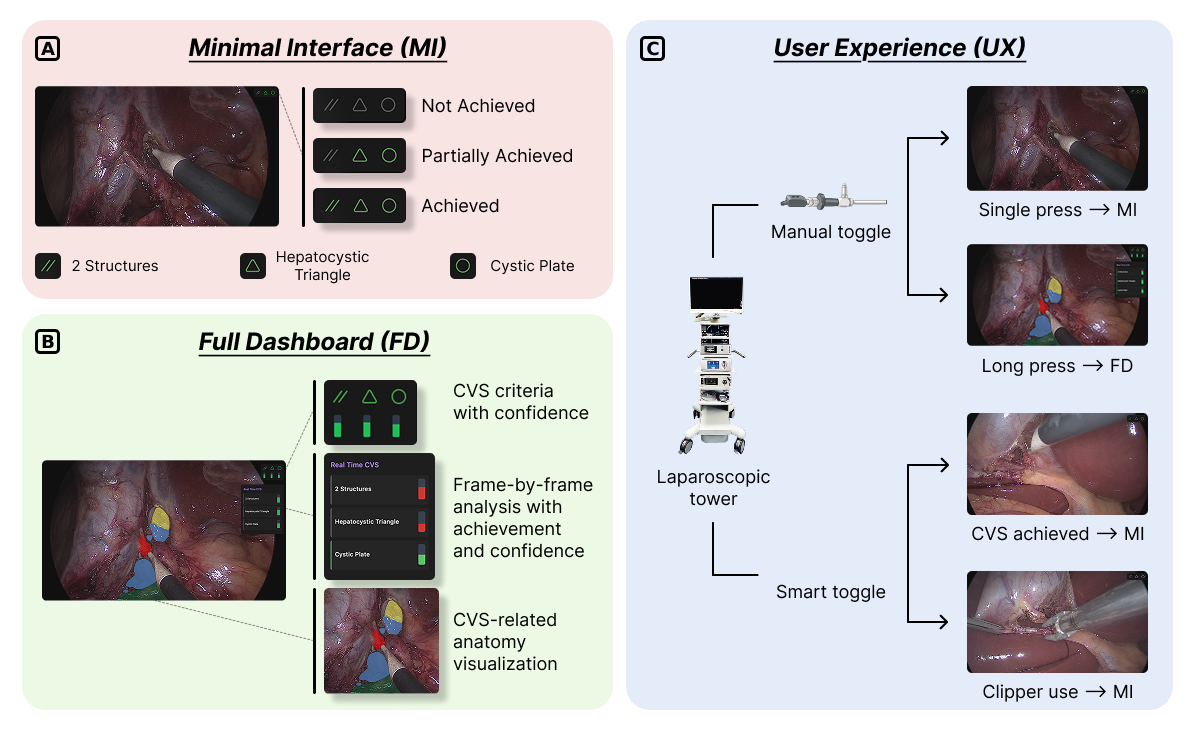}
    \caption{\textbf{CVS Copilot interface.} \\(A) Minimal Interface (MI) displaying the status of the three Critical View of Safety (CVS) criteria through a compact corner indicator, where for each criterion, gray denotes not achieved and green denotes achieved. (B) Full Dashboard (FD) providing detailed information including confidence scores, frame-level CVS assessment, and anatomy overlays, with the cystic duct shown in green, cystic artery in red, hepatocystic triangle in blue, and cystic plate in yellow. (C) User experience workflow showing manual and AI-triggered mechanisms that activate the CVS Copilot and how to switch between MI and FD intraoperatively.}
    \label{fig:fig2}
\end{figure*}

\section{Discussion}
\label{sec:discussion}

This study explored surgeons’ preferences regarding the design and integration of an intraoperative interface for AI-based CVS assessment. Across interviews, surgeons consistently favored interfaces that were visual, minimally intrusive, surgeon-controlled, and integrated within existing operative workflows. Participants also emphasized the importance of preserving surgeon autonomy, limiting unnecessary cognitive load, and adapting information delivery according to clinical context and level of expertise. These findings informed the design of a dual-mode HMI concept combining a minimal overlay with an optional expanded dashboard for on-demand interpretation.
A key contribution of this work is the identification of a narrow range of interface characteristics that surgeons consider acceptable during live surgery, which are summarized in Figure~\ref{fig:fig2}. Persistent overlays, unsolicited alerts, verbose explanations, and additional hardware were consistently rejected as sources of distraction and increased cognitive load, particularly when they required attention during high-demand moments. These findings have direct implications for how trust in intraoperative AI systems is established. Contrary to assumptions that trust increases with greater transparency or continuous feedback, surgeons associated trust with restraint, predictability, and respect for clinical judgment. Similar dynamics have been described in other clinical AI domains, where systems perceived as prescriptive or misaligned with practice are frequently overridden or ignored despite high accuracy~\cite{bussone_role_2015,tonekaboni_what_2019}.

The study also challenges dominant paradigms of explainable AI derived from perioperative and diagnostic decision-support systems, where clinicians often request access to contributing variables, feature importance, or model rationale to support deliberative decision-making~\cite{amann_explainability_2020,davidson_human-centered_2025}.In contrast, surgeons in this study articulated different expectations for intraoperative explainability. Rather than favoring textual or numerical rationales delivered continuously, they preferred explanations conveyed through spatially localized visual cues that directly indicate where attention or further dissection is required. When additional information was requested, causal clarification regarding unmet CVS criteria was sometimes sought, but this was considered secondary to spatial guidance during active dissection. These findings align with prior work in surgical ergonomics, where cluttered or persistent overlays are rejected, reinforcing the need for workflow-integrated, role-adaptive, and layered designs~\cite{liu_advances_2024,wong_cognitive_2024}.

Timing of feedback emerged as a critical determinant of acceptability. Surgeons consistently identified key operative decision points, such as the division of the cystic duct and artery during laparoscopic cholecystectomy, as the moment when AI assistance is most valuable, consistent with the original intent of CVS as a safety checkpoint. Continuous prompting throughout the procedure was viewed as unnecessary and potentially disruptive. Earlier guidance was tolerated only when clearly supportive and easily dismissible. These preferences align with prior reports on alarm fatigue and real-time clinical monitoring, where excessive or poorly timed alerts reduce vigilance and undermine trust~\cite{sendelbach_alarm_2013,cvach_monitor_2012}. Similar conclusions have been reached in studies of intraoperative augmented visualization, reinforcing the need for temporally sparse and context-aware systems~\cite{meola_augmented_2017}.

The dual-mode interface concept developed in this study represents a practical design response to these constraints. A minimal, surgeon-controlled interface allows background monitoring without occluding the operative field, while an expanded, on-demand dashboard provides additional detail when uncertainty arises or when educational value is sought. Layered interaction strategies of this kind are well established in other safety-critical domains, such as anesthesia and aviation, where escalation pathways support deeper inspection without forcing continuous engagement~\cite{endsley_toward_1995}. Notably, surgeons in this study strongly rejected hard-stop logic or punitive alerting, emphasizing that such mechanisms are incompatible with complex operative scenarios and risk shifting responsibility away from the surgeon, an issue previously highlighted in discussions of automation bias and accountability in clinical AI~\cite{skitka_does_1999,sittig_new_2010}.

Differences across experience levels further support the need for adaptive, dual-mode interfaces rather than static designs. Senior surgeons favored near-binary confirmation at decisive moments, whereas residents valued earlier guidance and richer contextual feedback during dissection. However, both groups agreed that increased detail should never be imposed by default. Similar role-dependent preferences have been reported in XAI research, where trainees and attendings differ in their informational needs and tolerance for detail~\cite{tonekaboni_what_2019}. The findings suggest that role-aware defaults can be implemented through controlled disclosure of information within a single system, rather than through separate interfaces.

Taken together, these results highlight the importance of interface design for the successful clinical integration of intraoperative AI. Even highly accurate predictions may be rejected if delivered at the wrong time, in an inappropriate form, or in a way that undermines surgical autonomy. The design insights derived from the development of CVS Copilot suggest a set of principles for intraoperative AI: minimal by default, expandable on demand, spatially grounded, temporally constrained, and explicitly non-coercive. While developed in the context of laparoscopic cholecystectomy, these principles are likely applicable to a broader class of intraoperative AI systems operating in safety-critical environments.
Several limitations should be acknowledged. First, the interviewers’ background in surgical AI research may have influenced question framing and surgeon responses, introducing the possibility of social desirability bias (i.e., participants may have expressed more favorable attitudes toward intraoperative AI assistance due to awareness of the interviewers’ research interests). While reflexive practices and peer debriefing were employed throughout the analytic process, complete neutrality cannot be guaranteed. Second, although the sample size was sufficient to achieve thematic saturation for the study aims, it limits the robustness of subgroup comparisons, particularly across levels of surgical experience. Third, data collection relied on remote interviews and static interface mock-ups rather than live intraoperative use or high-fidelity simulation, which may not fully capture attentional demands, stress, time pressure, and multitasking inherent to the operating room environment. Fourth, this study focused on perceived acceptability and design preferences rather than clinical effectiveness. As a result, no conclusions can be drawn regarding the impact of the proposed interface on real-world acceptance and usability, let alone CVS achievement, operative decision-making, or patient outcomes.
Future work should therefore prioritize the implementation of a functional CVS Copilot prototype and its evaluation in progressively realistic settings. Initial assessments may include video-based decision tasks and simulation-based studies designed to quantify workload, attention, decision confidence, and error patterns under controlled conditions. Subsequent in situ evaluations in the operating room will be required to assess real-world usability, integration into surgical workflow, and tolerance to model uncertainty and failure modes. Ultimately, prospective clinical studies will be necessary to determine whether surgeon-centered interface design for AI-based CVS assessment translates into measurable improvements in safety-related outcomes.

\section{Methods}
\label{sec:methods}

\begin{figure*}[t]
    \centering
    \includegraphics[width=1\linewidth]{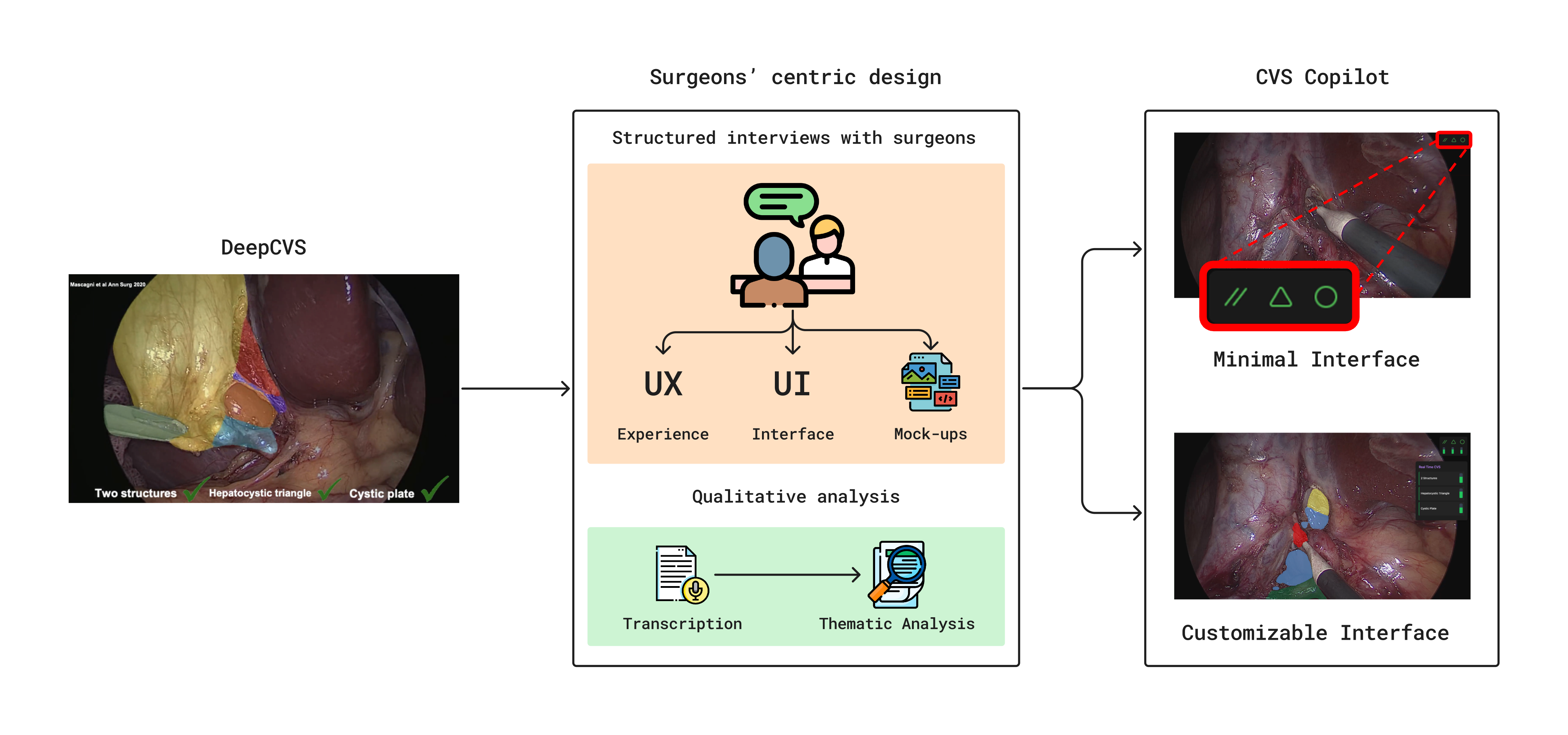}
    \caption{\textbf{Surgeon-centric design process.} \\ DeepCVS outputs informed structured interviews on User Experience (UX), User Interface (UI), and mock-ups, followed by qualitative analysis, leading to CVS Copilot (minimal overlay and customizable dashboard).}
    \label{fig:fig3}
\end{figure*}

\subsection{Study design}

A qualitative-dominant, mixed-methods study grounded in a UCD framework was conducted to elicit surgeons’ requirements for an intraoperative interface presenting AI predictions for CVS assessment in LC (see Figure~\ref{fig:fig3}. In this design, qualitative data served as the primary source of evidence through in-depth interviews and thematic analysis, while quantitative summaries were used to support and contextualize the qualitative findings. This study was reported in accordance with relevant elements of the COREQ checklist (Supplementary Table 1)~\cite{tong_consolidated_2007}. The study received an institutional review board waiver. All participants provided written informed consent.

\subsection{Population}

Eligible participants were surgeons or senior general surgery residents with LC experience. Purposive sampling was used to ensure variation in clinical role and seniority (residents, staff surgeons, professors) across institutions in Europe and the United States. Surgeons were contacted via professional and academic networks, informed of the study aims and interviewers’ backgrounds, and provided consent for recording. Recruitment continued until thematic saturation was reached~\cite{guest_simple_2020}. Some surgeons were known to the interviewers prior to the study; no prior supervisory or evaluative relationships existed.

\subsection{Interviews}

Semi-structured interviews were conducted remotely via Google Meet videoconference over two weeks in February 2025, each scheduled for approximately 60 minutes. Audio-video was recorded using OBS Studio.  All interviews were conducted by NC and LA, a medical student and a physician-researcher, respectively. Both are male and were trained in qualitative research methods prior to the study. Three non-participants (research team members) were present during one interview for observation purposes but did not contribute. Notes were not collected during the interviews; instead, interpretations and decisions were documented throughout the analysis to ensure consistency in theme development. Transcripts were not returned to surgeons for comment or correction.  
The interview guide was developed in three steps. First, an in-depth review of outputs from an existing AI system for CVS assessment (DeepCVS~\cite{mascagni_artificial_2022}) was performed to anchor questions in realistic model capabilities and failure modes. Second, interface mock-ups were created in Microsoft PowerPoint to serve as concrete discussion probes. Third, wording and clinical relevance were iteratively refined with a practicing surgeon outside the study group.
Interviews addressed three domains: UX, UI, and mock-up evaluation. Within the UX domain, discussions explored attitudes toward intraoperative AI, information needs for CVS confirmation, timing and triggering of assistance, interaction and control, notification channels, and the tools on which AI outputs should appear. To surface role-dependent needs, surgeons were asked to provide perspectives from three roles: department chief, staff surgeon, and resident. The UI domain focused on visualization strategies, including preferences for displaying AI-predicted CVS-related information, the level of detail considered appropriate during surgery, the types of predictions regarded as trustworthy, and broader design considerations. The mock-up evaluation domain presented surgeons with two scenarios: (I) a minimal overlay integrated on the main laparoscopic feed and (II) an expanded, on-demand dashboard. Across these scenarios, systematic variation was introduced in activation mode (manual toggle versus persistent display), feedback channel (iconic/visual cues versus short text versus voice), information format (binary indicators versus percentages/continuous scales), and annotation scope (selective highlighting of critical anatomy versus broader overlays). Following open discussion, surgeons provided structured preferences (e.g., accept/reject or ordinal ratings) for each alternative. Table~\ref{tab:table3} reports the full list of questions.

\begin{table*}[t]
{\small
\setlength{\tabcolsep}{3pt}
\renewcommand{\arraystretch}{1.1}
\centering
\caption{Interview guide domains and example questions covering background attitudes, user experience (UX), user interface (UI), and preference for mock-ups.}
\label{tab:table3}
\begin{tabular}{p{0.2\textwidth} p{0.72\textwidth}}
\toprule
\textbf{Category} & \textbf{Questions} \\
\midrule
Background &
Would you like AI assistance in the Operating Room? Why / Why not? \newline
How do you envision the interplay between the AI system and the surgeons? \\
\addlinespace
User Experience (UX) &
How would you like AI specifically to assist with regard to CVS? \newline
At what point in the procedure should AI help confirm CVS has been achieved? \newline
Would notifications be useful during surgery? \newline
Would you find interactive features useful? \newline
How should AI integrate into the surgical workflow to minimize disruption? \newline
What tools should AI data appear on? \\
\addlinespace
User Interface (UI) &
How should the platform visualize AI-predicted CVS-related information? \newline
Would you prefer to see detailed explanations during the procedure? \newline
What kinds of predictions would you trust AI to provide? \newline
What type of design choices do you think are appropriate? \\
\addlinespace
Mock-ups &
Please describe the elements that you like and those that you would improve in the minimal overlay. \newline
Please describe the elements that you like and those that you would improve in the dashboard. \\
\bottomrule
\end{tabular}
}
\end{table*}

\subsection{Analysis}

Recordings were automatically transcribed with a custom Python (version 3.12.11) script using Whisper AI and then manually verified for accuracy. Transcripts were anonymized and stored on an encrypted institutional server. Qualitative analysis followed Braun and Clarke’s six-phase reflexive thematic  analysis~\cite{braun_toward_2023}. Transcripts were segmented into discrete excerpts (i.e., meaningful portions of text), which were assigned one or more thematic codes during analysis. Two researchers independently coded an initial subset of transcripts to align on a shared codebook, then continued coding reflexively with regular peer debriefs. In a final mapping step, themes and representative codes were aligned to human-factors heuristics~\cite{nielsen_heuristic_1990} (standard design principles commonly used to assess whether an interface presents information clearly, minimizes cognitive load, and preserves user control) to derive interface requirements. Quantitative analyses were used to describe surgeons’ characteristics and compare structured preference responses across experience strata. Descriptive summaries of surgeons’ characteristics and preference responses were performed in Microsoft Excel. An audit trail documented protocol decisions, codebook revisions, and versioned mock-up iterations.

\section*{Competing Interests}

Authors PM and NP are co-founders and shareholders of Scialytics. All other authors have no conflicts of interest to report.

\section*{Data Availability}

The datasets generated and/or analyzed during the current study are not publicly available due to the qualitative nature of the interviews and the potential identifiability of participants, but are available from the corresponding author on reasonable request.

\section*{Code Availability}

The underlying code of the CVS Copilot is not publicly available but may be made available to qualified researchers on reasonable request from the corresponding author.

\section*{Acknowledgments}

We thank the surgeons who were interviewed, listed in alphabetical order: Alberto Arezzo, Ludovica Baldari, Andrea Balla, Francesco Brucchi, Ivan Capobianco, Claudio Fiorillo, Alain Garcia, Federico Gheza, Daniel N. Hashimoto, Giovanni Laracca, Giuseppe Massimiani, Salvador Morales Conde, Vinicio Mosca, Monica Ortenzi, Andrea Spota, Francesco Taliente, and Maria Vannucci.

\section*{Funding Statement}

This work was developed within the Interdisciplinary Thematic Institute HealthTech (ITI 2021--2028 program of the University of Strasbourg, CNRS and Inserm), supported by IdEx Unistra (ANR-10-IDEX-0002) and SFRI (STRATUS project, ANR-20-SFRI-0012) under the framework of the French Investments for the Future Program.

This work has received funding from the European Union (ERC, CompSURG, 101088553). Views and opinions expressed are, however, those of the authors only and do not necessarily reflect those of the European Union or the European Research Council. Neither the European Union nor the granting authority can be held responsible for them. This work was also partially supported by French state funds managed by the ANR under Grant ANR-10-IAHU-02.

\section*{CRediT Authorship Contribution Statement}

LA: Conceptualization, Methodology, Validation, Formal analysis, Investigation, Data curation, Writing -- Original Draft, Writing -- Review \& Editing, Visualization.

NC: Methodology, Validation, Formal analysis, Investigation, Data curation, Writing -- Review \& Editing.

AM: Methodology, Writing -- Review \& Editing.

PM: Conceptualization, Methodology, Writing -- Review \& Editing, Supervision.

NP: Conceptualization, Writing -- Review \& Editing, Supervision, Resources, Funding Acquisition, Project Administration.

\bibliographystyle{splncs04}
\bibliography{arxiv}

\clearpage

\vspace*{\fill}
\begin{center}
{\LARGE\bfseries Supplementary Materials}

\vspace{2cm}

\noindent
\textbf{Supplementary Video 1}

\vspace{0.5cm}

\href{https://seafile.unistra.fr/f/2e94fdb2a3504c0e9caf/}
{\Large\texttt{View Supplementary Video 1}}

\vspace{0.8cm}

\qrcode[height=3cm]{https://seafile.unistra.fr/f/2e94fdb2a3504c0e9caf/}

\vspace{0.5cm}

{\small Scan the QR code or click the link above to access the video.}
\end{center}
\vspace*{\fill}

\clearpage



\section*{Supplementary material}

# Supplementary Materials

**Supplementary Video 1**

`View Supplementary Video 1`

Scan the QR code or click the link above to access the video.

# COREQ (COnsolidated criteria for REporting Qualitative research) Checklist



A checklist of items that should be included in reports of qualitative research. You must report the page number in your manuscript where you consider each of the items listed in this checklist. If you have not included this information, either revise your manuscript accordingly before submitting or note N/A.

| Topic | Item No. | Guide Questions/Description | Reported on Page No. |
|---|---|---|---|
| **Domain 1: Research team and reflexivity** | | | |
| *Personal characteristics* | | | |
| Interviewer/facilitator | 1 | Which author/s conducted the interview or focus group? | 7 |
| Credentials | 2 | What were the researcher's credentials? E.g. PhD, MD | 7 |
| Occupation | 3 | What was their occupation at the time of the study? | 7 |
| Gender | 4 | Was the researcher male or female? | 7 |
| Experience and training | 5 | What experience or training did the researcher have? | 7 |
| *Relationship with participants* | | | |
| Relationship established | 6 | Was a relationship established prior to study commencement? | 7 |
| Participant knowledge of the interviewer | 7 | What did the participants know about the researcher? e.g. personal goals, reasons for doing the research | 7 |
| Interviewer characteristics | 8 | What characteristics were reported about the inter viewer/facilitator? e.g. Bias, assumptions, reasons and interests in the research topic | 7 |
| **Domain 2: Study design** | | | |
| *Theoretical framework* | | | |
| Methodological orientation and Theory | 9 | What methodological orientation was stated to underpin the study? e.g. grounded theory, discourse analysis, ethnography, phenomenology, content analysis | 7,8 |
| *Participant selection* | | | |
| Sampling | 10 | How were participants selected? e.g. purposive, convenience, consecutive, snowball | 7 |
| Method of approach | 11 | How were participants approached? e.g. face-to-face, telephone, mail, email | 7 |
| Sample size | 12 | How many participants were in the study? | 2,7 |
| Non-participation | 13 | How many people refused to participate or dropped out? Reasons? | N/A |
| *Setting* | | | |
| Setting of data collection | 14 | Where was the data collected? e.g. home, clinic, workplace | 7 |
| Presence of non-participants | 15 | Was anyone else present besides the participants and researchers? | 7 |
| Description of sample | 16 | What are the important characteristics of the sample? e.g. demographic data, date | 2 |
| *Data collection* | | | |
| Interview guide | 17 | Were questions, prompts, guides provided by the authors? Was it pilot tested? | 7,8 |
| Repeat interviews | 18 | Were repeat inter views carried out? If yes, how many? | N/A |
| Audio/visual recording | 19 | Did the research use audio or visual recording to collect the data? | 7 |
| Field notes | 20 | Were field notes made during and/or after the inter view or focus group? | 7 |
| Duration | 21 | What was the duration of the inter views or focus group? | 7 |
| Data saturation | 22 | Was data saturation discussed? | 2,7 |
| Transcripts returned | 23 | Were transcripts returned to participants for comment and/or | 7 |

| Topic | Item No. | Guide Questions/Description | Reported on Page No. |
|---|---|---|---|
| | | correction? | |
| **Domain 3: analysis and findings** | | | |
| *Data analysis* | | | |
| Number of data coders | 24 | How many data coders coded the data? | 8 |
| Description of the coding tree | 25 | Did authors provide a description of the coding tree? | 3,4,8 |
| Derivation of themes | 26 | Were themes identified in advance or derived from the data? | 2,8 |
| Software | 27 | What software, if applicable, was used to manage the data? | 8 |
| Participant checking | 28 | Did participants provide feedback on the findings? | 2 |
| *Reporting* | | | |
| Quotations presented | 29 | Were participant quotations presented to illustrate the themes/findings? Was each quotation identified? e.g. participant number | 3,4 |
| Data and findings consistent | 30 | Was there consistency between the data presented and the findings? | 2-6 |
| Clarity of major themes | 31 | Were major themes clearly presented in the findings? | 2-6 |
| Clarity of minor themes | 32 | Is there a description of diverse cases or discussion of minor themes? | 2-6 |

Developed from: Tong A, Sainsbury P, Craig J. Consolidated criteria for reporting qualitative research (COREQ): a 32-item checklist for interviews and focus groups. *International Journal for Quality in Health Care*. 2007. Volume 19, Number 6: pp. 349 – 357

**Once you have completed this checklist, please save a copy and upload it as part of your submission. DO NOT include this checklist as part of the main manuscript document. It must be uploaded as a separate file.**



\end{document}